\documentclass[twocolumn,aps,prl,preprintnumbers,showpacs,nofootinbib]{revtex4}
\usepackage[utf8]{inputenc}
\usepackage[english]{babel}
\usepackage{amsmath}
\usepackage{amsfonts}
\usepackage{amssymb}
\usepackage{epsfig}
\usepackage{graphics,psfrag,rotating}
\usepackage{graphicx}
\usepackage{dcolumn}
\usepackage{bm}
\usepackage{epstopdf}
\usepackage{color}
\usepackage[usenames,dvipsnames,svgnames]{xcolor}
\usepackage[colorlinks=true,
            linkcolor=red,
            urlcolor=gray,
            citecolor=blue]{hyperref}
  \usepackage{hyperref}

\newcommand{\lsim}   {\mathrel{\mathop{\kern 0pt \rlap
{\raise.2ex\hbox{$<$}}}
 \lower.9ex\hbox{\kern-.190em $\sim$}}}
\newcommand{\gsim}   {\mathrel{\mathop{\kern 0pt \rlap
{\raise.2ex\hbox{$>$}}}
\lower.9ex\hbox{\kern-.190em $\sim$}}}

\def\3nab{\tilde{\nabla}}

\def\hsp5{\hspace{5mm}}

\def\case#1/#2{\textstyle\frac{#1}{#2}}

\newcommand{\ber}{\begin{equation}}
\newcommand{\eer}{\end{equation}}
\newcommand{\baa}{\begin{eqnarray}}
\newcommand{\eaa}{\end{eqnarray}}

\def\bc {\begin{center}}
\def\ec {\end{center}}
\def\case#1/#2{\frac{#1}{#2}}

\newcommand{\ew}{\end{widetext}}

\newcommand{\bee}{\begin{equation}}
\newcommand{\bse}{\begin{subequation}}
\newcommand{\ese}{\end{subequation}}
\newcommand{\ee}{\end{equation}}
\newcommand{\eei}{\end{eqnarray}\indent\indent}
\newcommand{\ba}{\begin{array}}
\newcommand{\ea}{\end{array}}
\newcommand{\bal}{\begin{eqnarray}}
\newcommand{\eal}{\end{eqnarray}}

\def\case#1/#2{\textstyle\frac{#1}{#2} }

\newcommand{\bver}{\begin{verbatim}}
\newcommand{\ever}{\end{verbatim}}

\begin{document}

\title{Brout-Englert-Higgs mechanism for accelerating observers}


\author{ Antonio Dobado \\
Departamento de  F\'\i sica Te\'orica I\\
 Universidad Complutense, 28040 Madrid, Spain.
}

\date{\today}

\begin{abstract}

In this work we consider the spontaneous symmetry breaking of the electroweak $SU(2)_L\times U(1)_Y$ gauge group into $U(1)_{em}$ taken place in the  Standard Model of particle physics as seen from the point of view of an accelerating observer. According to the Unruh effect that observer detects the Minkowski vacuum as a thermal bath at a temperature proportional to the proper acceleration $a$. Then we show that (in a certain large $N$ limit) when the acceleration is bigger than the critical value $a_c = 4 \pi v$ (where $v$ is the Higgs vacuum expectation value), the electroweak  $SU(2)_L\times U(1)_Y$ gauge symmetry is restored and all elementary particles become massless. In addition, even  observers with $a<a_c$, can see this symmetry restoration in the region close to the Rindler horizon.

\end{abstract}
\maketitle

\vspace{0.1cm}

\section{Introduction}

In the present day, our positive  knowledge about fundamental interactions can be summarized in just two theories. On the one hand we have the Standard Model (SM) of particle physics and on the other hand we have General Relativity (GR). The SM is a Quantum Field Theory (QFT) invariant under the (chiral) gauge group $SU(3)_C\times SU(2)_L \times U(1)_Y$ which describes strong and electroweak interactions between elementary particles (leptons and quarks). GR is a classical theory of gravity incorporating the Equivalence Principle  and the curvature of space-time as essential ingredients. 

In addition the SM provides a mechanism for the generation of the masses of the elementary particles (not for the composite ones as the proton). This is the celebrated Brout-Englert-Higgs (BEH) mechanism \cite{EB,H} (or Higgs mechanism for short) which seems to be strongly supported by the discovery in 2012 at the CERN Large Hadron Collider (LHC) of a particle with properties  compatible with those expected for the Higgs boson. 

Therefore, in spite of the many well known problems still to  be solved, such as the problem of Dark Matter (DM),  Dark Energy (DE),  baryogenesis, strong CP, neutrino masses and many others, a great deal of   
observed phenomena  can in principle be accommodated in the SM formulated
in a curved space-time background. Also it is generally believed that a better understanding of QFT in curved backgrounds could deliver a deepest 
insight on the fusion of GR and Quantum Mechanics  (QM) as the two pillars of modern theoretical physics.

The formulation of QFT for arbitrary observers, or in the presence of gravitational fields, is non-trivial mainly due to the possible presence of horizons, the best known examples of this being the Hawking radiation \cite{Hawking} and the  Unruh effect \cite{Unruh} (see \cite{Crispino:2007eb} for a very complete review). In this note we will concentrate in the second one and in its connection with the Higgs mechanism. As it is well known, trying to understand better the Hawking radiation, Unruh realized that an observer moving through the Minkowski vacuum 
with a constant acceleration $a$ will detect a thermal bath at temperature:
\ber
T=\frac{a \hbar}{2\pi c k_B}.
\eer

This result can be obtained and confirmed in different ways. Operationally by studying the response of a so-called  Unruh-DeWitt detector to the quantum fluctuations of the fields. For the free field case one can also use Bogolyubov transformations which  was the approach used by the pioneers of field quantization on Rindler space \cite {FullingBirrelParkerBoulware}. Also it is possible to consider operator algebra  (see \cite{Haag:1992hx} ) in the context of 
Modular Theory where the concept  of  KMS (Kubo-Martin-Schwinger \cite{KMS})  states plays an essential role (see   \cite{Earman:2011zz} ).

Notice that the formula above relates QM, Relativity and Statistical Physics since it contains the Planck constant $\hbar$, the speed of light $c$ and the Boltzmann constant $k_B$ (in the following we will use natural units with $c= \hbar = k_B= 1$). That shows its fundamental nature in spite of the difficulties for its experimental confirmation. For example Bell and Leinaas have suggested the possibility of observing the Unruh effect by measuring the polarization of electrons in storage rings \cite{Bell} but  that interesting possibility is still under discussion \cite{Crispino:2007eb}.

Our results in this work will be based in the so-called Thermalization Theorem. It was introduced by Lee \cite{Lee} and it consists in a  path integral approach to QFT for arbitrary observers and curved space time. It can be applied to any kind of field appearing in the SM (scalars, fermions or gauge bosons) and most importantly, to interacting systems. Moreover, the result does not rely on perturbation theory or any other particular treatment of the interaction.
This means that the Unruh effect could give rise to non-trivial dynamical effects such as phase transitions. Indeed it has been shown that accelerating observes do observe a restoration of continuous global symmetries in some systems featuring Spontaneous Symmetry Breaking (SSB). For example in the Nambu-Jona-Lasinio model \cite{Ohsaku:2004rv},  
 in  the $\lambda \Phi^4$ theory at the one-loop level \cite{Castorina:2012yg} and in the  Linear Sigma Model (LSM) in the large $N$ limit \cite{Dobado}.

Therefore it seems natural to wonder if this restoration of symmetry due to the Unruh effect applies to gauge symmetries too. For this reason we will consider in this work the Higgs mechanism of the SM as seen by an accelerating observer. In the SM the electroweak $SU(2)_L \times U(1)_Y$ gauge symmetry is spontaneously broken by the Higgs sector to the electromagnetic gauge group $U(1)_{em}$. As a consequence of that the three Goldstone bosons corresponding to the broken generators are eaten by a combination of the gauge boson to give masses to the $W^{\pm}$ and $Z$ electroweak bosons leaving the $A$ photon field massless (the BEH or Higgs mechanism). The Higgs system is just a  complex doublet featuring a global $SU(2)_L\times SU(2)_R$ global symmetry. A potential is introduced {\it  ad hoc } to produce a SSB of this symmetry down to $SU(2)_{L+R}$. When the Higgs sector is coupled with the   $SU(2)_L \times U(1)_Y$ gauge fields, this global SSB triggers the Higgs mechanism. The Higgs sector can alternatively be described by a real  four-multiplet with global symmetry $SO(4)$ spontaneously broken to $SO(3)$. Thus the first three fields are the would-be Goldstone bosons and the fourth corresponds to the Higgs boson.  

In this work we will
study the SSB of the SM gauge symmetry for accelerating observers by using the Thermalization Theorem and the large $N$ limit. This approximation is a non-perturbative way for computing non-trivial dynamical effects, such as phase transitions, 
which is quite convenient for our purposes here.  In order to implement it, we will generalize the SSB pattern of the Higgs system to $SO(N+1)$ down to $SO(N)$, then we will do the relevant computations to the leading order in the $1/N$ expansion and finally we will set $N=3$ again.     

As mentioned above the  $SU(2)_L\times U(1)_Y$ gauged   Higgs system  features a SSB of the gauge symmetry down to $U(1)_{em}$. However, at higher temperatures, the system experiments a thermal second order phase transition corresponding to a $SU(2)_L\times U(1)_Y$ symmetry restoration at a temperature $T_c=2v $ in the large $N$ limit, with  $v \simeq 245$ GeV being the Higgs vacuum expectation value (VEV). In this work we will show that a similar phase transition is experimented by an accelerating observer with constant acceleration $a$ at the critical acceleration $ a_c = 4 \pi v \simeq 3$ TeV  as computed in the large $N$ approximation. As a consequence the electroweak $W^{\pm}$ and $Z$ bosons become massless for such an observer. Moreover, as we will see below, even if the acceleration $a$ is smaller than the critical one $a_c$, the accelerating observer will perceive  a restoration of the gauge symmetry in the region close to her (Rindler) horizon. This is due to the fact that Rindler space is not homogeneous  and this  produces the interesting effect of having a Higgs VEV which  is position dependent. Thus the observer sees the symmetry broken when looking in the direction of the acceleration but she observes a restoration of the symmetry somewhere in the opposite direction. 

Now one may wonder if there is any possible physical scenario where the effect described in this work could have any relevance. In \cite{Kharzeev:2005iz}  and \cite{DiasdeDeus:2006xk} the authors introduced a model for hadron thermalization in Heavy Ion Collisions (HIC) based on the Unruh effect which could be applied to the description of BNL Relativistic Heavy Ion Collider (RHIC) results then available. The corresponding Unruh temperature in this case is about $175$ MeV corresponding to the chiral or deconfinement phase transition at an acceleration of the order of one GeV. Currently the LHC is producing proton-proton collisions at a center of mass energy of $13$ TeV which corresponds typically to parton-parton interactions at several TeV's of center of mass energy. Therefore it is not unthinkable to envision  the possibility of electroweak symmetry restoration by acceleration playing a role at the LHC. In any case this requires a detailed analysis of this physical case which is far beyond the scope of this work.

This paper is organized as follows; in section II we define the Rindler and comoving coordinates in Rindler space and we enunciate the Thermalization Theorem to be used later.  In section III we introduce the large $N$ limit of the Higgs sector of the SM considered in this work and we compute in this  limit the partition function relevant for the Thermalization Theorem. Section IV is dedicated to the electroweak symmetry restoration obtained and the details of the VEV profile for the accelerating observers. In section V we comment on different aspects of our results and section VI is dedicated to the conclusions. Finally Appendices A and B are devoted to the mathematical details of the computations needed for this work.   

\section{Comoving coordinates and the Thermalization Theorem}
 
In order to describe how it is possible to obtain the above results we start from the Minkowski space  metric written in terms of Cartesian (inertial) coordinates  $X^\mu =(T,X,Y,Z)$:
\ber
ds^2= dT^2-dX^2-dX_\bot^2
\eer
where $X_\bot = (Y,Z)$.
 Dealing with accelerating observers (or detectors) in Minkowski space it is very useful to consider Rindler and comoving coordinates. 
 Rindler coordinates are defined as:
\baa
T & = & \rho\, \sinh\, \eta   \nonumber \\       
X   &  = & \rho\, \cosh\, \eta        \label{Rindler}
\eaa
where $\rho \in (0,\infty)$ and $\eta \in (-\infty,\infty)$. As it is well known these coordinates cover only 
 the region $X > 
\mid T\mid$ (the ${\cal R}$ wedge). Similar coordinates can be introduced covering the left wedge ${\cal L}$ where $- X > \mid T \mid$.
In the  ${\cal R}$  region the metric reads:
\ber
ds^2= \rho^2d\eta^2- d\rho^2-dX_\bot^2.
\eer
The two other regions are the origin past ${\cal P}\   (T<- \mid X\mid) $ and the origin future  ${\cal F}\  (T> \mid X\mid) $. An uniformly accelerating observer in the
$X$  direction (and constant  $X_\bot$)  with proper acceleration $a$  will follow a world line 
described in Rindler coordinates by the simple equations:
  $\rho   =  1/a$ and   
$ \eta   =  a \tau$ with $\tau$ being the proper time.
Therefore Rindler coordinates correspond to a network of
observers with different proper constant acceleration $a= 1/
\rho$ and having a clock measuring their proper times in units
of $1/a$. The important thing for our work here is that those observers have a past and a future horizon at $X=-T$ and
$X=T$ respectively which they find in the infinite remote past  or
future (in proper time)  or also in the limit $\rho
\rightarrow 0$ (infinite acceleration).

It is also interesting to introduce on ${\cal R}$ the coordinates $x^\mu=(t,x,y,z)$ defined as:
\baa
T & = & \frac{1}{a}e^{a x}  \sinh\,(a t)  \nonumber  \\
X   &=   &\frac{1}{a}e^{a x}  \cosh\,(a t)  \nonumber  \\
Y  & =  &y    \nonumber  \\
Z  & =  & z.        \label{Comoving}
\eaa
These are the comoving coordinates associated to some particular non-rotating
accelerating observer located at $x=0$ with constant acceleration $a$ in the $X$
direction. Note that $t
,x,y,z \in (-\infty, \infty)$ and one has $\rho= e^{a x}/a$ and
$\eta= a t$. In these coordinates the metric reads:
\ber
ds^2 = e^{2 a x}(dt^2-dx^2)-dx_\bot^2
\eer
where $t$ is the observer's (located at $x=0$) proper time and $x_\bot=(y,z)$. In the limit of vanishing $a$ we recover Minkowski metric as it must be.

The Thermalization Theorem \cite{Lee}, giving rise to the Unruh effect stems from the following essential fact:  an accelerating observer can only feel directly the Minkowski vacuum fluctuations inside ${\cal R}$. However those fluctuations are entangled with the ones corresponding to the left Rindler region  ${\cal L}$ ($    X < -\mid T \mid  $). As a consequence of that she will see the Minkowski vacuum (by that we mean the true ground state of the system including interactions)
 as a mixed state described by a density matrix $\rho _R$ which, according to the Thermalization Theorem \cite{Lee}, can be written in terms of the Rindler Hamiltonian $\hat H_R $ (the generator of the $t$ time translations) as:  
\ber
\hat \rho_R = \frac{ e^{-2\pi \hat H_R/a}    }{Tr (e^{-2\pi \hat H_R/a})}.
\eer
In particular the expectation  value of an operator $\hat A_R$ defined on the Hilbert space ${\cal H}_R$ corresponding to the region ${\cal R}$ in the Minkowski vacuum $\mid \Omega_M>$ is given by:
\ber
< \Omega_M \mid  \hat A_R \mid \Omega_M > = Tr  (\hat \rho_R \hat A_R).
\eer
This is just what one would find  in a  thermal ensemble at temperature $T= a/ 2 \pi$ (in natural units) and it can be understood as a very precise formulation of the Unruh effect. 

\section{The large $N$ limit of the SM Higgs sector in Rindler space}

Now one can try to apply this result to the case of the SM, in particular to its symmetry breaking sector. Thus, in order to study the Higgs mechanism for accelerating observers we consider the   $SU(2)_L \times U(1)_Y$
 gauged $SO(N+1 )/SO(N)$ linear sigma model defined by the Minkowski-space Lagrangian:
\ber
\mathcal{L}=\mathcal{L}_0 +\mathcal{L}_{YM}
\eer
where
\ber \label{eq:gaugedlsmlagrangian}
\mathcal{L}_0= \frac{1}{2} (D_{\mu} \Phi)^T  D^{\mu} \Phi -V
\left(\Phi^T \Phi
 \right).
\eer
The multiplet $\Phi^T=(\bar \pi,\sigma)$  contains $N+1$ real 
scalar fields  ($\bar \pi$  is a $N$ component scalar multiplet). The potential is given by:
\ber
V \left(\Phi^T \Phi
 \right)=
-\mu^2 \Phi^T \Phi + \lambda \left(\Phi^T \Phi
 \right)^2
\eer
where $\lambda$ is positive in order to have a potential
bounded from below and  $\mu^2$ is positive  
in order to produce  the SSB $SO(N+1) \rightarrow SO(N)$ . 
The covariant derivative is defined by:
\ber
D_{\mu} \Phi =(\partial_{\mu}   +V_\mu )\Phi= (\partial_{\mu}    - i g T^k_L W^k_{\mu}  -i g' T_Y B_{\mu})\Phi
\eer
where $W_{\mu}^k$ and $B_\mu$ are the $SU(2)_L$ and $U(1)_Y$ gauge fields respectively and $g$ and $g'$ are the corresponding gauge couplings. These groups are contained in $SO(N+1)$ (but not in $SO(N)$) for $N \ge 3$ and are generated here  by the $(N+1) \times (N+1)$ matrices $T_L^k = i M_L^k/2$ and $ T_Y = i M_Y / 2$ where:
\begin{equation}\label{ML1}
  M_L^1 = %
   \begin{pmatrix}
     0   &  0     & 0  &...     & - \\
     0   & 0     & -  &...      & 0    \\
      0        & +       & 0  & ... & 0 \\
        ...     &        &   &  &  \\
     + & 0     & 0 &...  & 0
   \end{pmatrix},
\end{equation}

\begin{equation}\label{ML2}
  M_L^2 = %
   \begin{pmatrix}
     0   &  0     & +  &...     & 0 \\
     0   & 0     & 0  &...      & -    \\
      -        & 0      & 0  &...  & 0 \\
        ...     &        &   &  &  \\
     0 & +     & 0 &...  & 0
   \end{pmatrix},
\end{equation}

\begin{equation}\label{ML3}
  M_L^3 = %
   \begin{pmatrix}
     0   &  -     & 0  &...     & 0 \\
     +   & 0     & 0  &...      & 0    \\
      0      & 0      & 0  &...  & - \\
        ...     &        &   &  &  \\
     0 & 0    & + &...  & 0
   \end{pmatrix}
\end{equation}
and
\begin{equation}\label{MY}
  M_Y = %
   \begin{pmatrix}
     0   &  -     & 0  &...     & 0 \\
     +   & 0     & 0  &...      & 0    \\
      0      & 0      & 0  &...  & + \\
        ...     &        &   & &  \\
     0 & 0    & - &...  & 0
   \end{pmatrix}.
\end{equation}
For example, for $N=4$ we have:
\begin{equation}\label{ML1}
  M_L^1 = %
   \begin{pmatrix}
     0   &  0     & 0  &0    & - \\
     0   & 0     & -  &0     & 0    \\
      0        & +       & 0  & 0 & 0 \\
        0     &     0   &  0 & 0 & 0 \\
     + & 0     & 0 &0 & 0
   \end{pmatrix},
\end{equation}

\begin{equation}\label{ML2}
  M_L^2 = %
   \begin{pmatrix}
     0   &  0     & +  &0     & 0 \\
     0   & 0     & 0  &0      & -    \\
      -        & 0      & 0  &0  & 0 \\
        0     &   0     &0   &0  &0  \\
     0 & +     & 0 &0 & 0
   \end{pmatrix},
\end{equation}

\begin{equation}\label{ML3}
  M_L^3 = %
   \begin{pmatrix}
     0   &  -     & 0  &0    & 0 \\
     +   & 0     & 0  &0      & 0    \\
      0      & 0      & 0  &0  & - \\
      0     &  0      & 0  & 0 & 0 \\
     0 & 0    & + &0  & 0
   \end{pmatrix}
\end{equation}
and
\begin{equation}\label{MY}
  M_Y = %
   \begin{pmatrix}
     0   &  -     & 0  &0     & 0 \\
     +   & 0     & 0  &0     & 0    \\
      0      & 0      & 0  &0  & + \\
        0    &  0      & 0  &0 & 0 \\
     0 & 0    & - &0  & 0
   \end{pmatrix}.
\end{equation}

Then it is easy to check $[T^i_L, T^j_L]= i \epsilon_{ijk}T_L^k$, $tr T_L^i T_L^j = \delta_{ij}$ and  $[T_L^k,T_Y]=0$. The Yang-Mills Lagrangian is defined as usual as:
\ber
\mathcal{L}_{YM} = -\frac{1}{4}W^i_{\mu\nu}W^{i\mu\nu} -\frac{1}{4}B_{\mu\nu}B^{\mu\nu}
\eer
with:
\ber
W^i_{\mu\nu}=\partial_\mu W_\nu^i-\partial_\nu W_\mu^i+g \epsilon_{ijk}W_\mu^j W_\nu^k
\eer
and
\ber
B_{\mu\nu}=\partial_\mu B_\nu-\partial_\nu B_\mu.
\eer
The SSB pattern induced by the potential is $SO(N+1)
\rightarrow SO(N)$ and it gives rise in principle to $N$ Goldstone bosons living in the coset space $S^N=SO(N+1)/SO(N)$. However, in this case the first three Goldstones are eaten by a particular combination of the gauge bosons which become massive (the Higgs mechanism). In particular the case $N=3$ corresponds exactly with the Yang-Mills plus Higgs sector of the SM and no Goldstone boson appear in the spectrum since all of then (three) are eaten to produce the masses for the $W_\mu^{\pm}$ and $Z_\mu$ electroweak bosons.
At  the tree level  the low-energy dynamics is controlled by the broken phase where:
\ber
< \Omega_M \mid  \hat{ \sigma} \mid \Omega_M> =v.
\eer
and  $v^2=\mu^2/2\lambda=NF^2$. Here we have introduced the constant $F$ to stress the fact that, as we will see below,  $v^2$ is order $N$ in the large $N$ limit considered here.

According to the Thermalization Theorem an accelerating observer will see the system  described by the above Lagrangian
as a canonical ensemble given  by the partition function:
\baa
Z_R(a) & = & Tr (e^{-\frac{2\pi}{a} \hat H_R}  )   \\
            &  =   & \int[dW][dB] [d\Phi] \exp
\left( -S_{RE}[\Phi,W,B] \right) \ ,    \nonumber  
\eaa
where $S_{RE}$ is the Euclidean action in Rindler space and the functional integrals are defined using  thermal-like  periodic boundary conditions. For example:
\ber
\Phi(\bar x, 0) = \Phi(\bar x,2\pi/a)
\eer
and also
\ber
\Phi(\mid \bar x \mid=\infty , t_E) ^T\Phi(\mid \bar x
\mid=\infty , t_E)= \sigma^2( \mid \bar x \mid=\infty , t_E) =
v^2, 
\eer
where $t_E$ is the Euclidean comoving time and $\bar x =(x,y,z)$.
In comoving coordinates the Euclidean action $S_{RE}[\Phi]$ defined on  ${\cal R}$  is:
\baa
S_{RE}[\bar \pi,\sigma,W,B] 
   & = &       
\int d^4x \sqrt{g}  [
\frac{1}{2}\Phi^T(-\square_E-4 \lambda v^2)\Phi    \nonumber   \\
& + &  \lambda (\Phi^T \Phi)^2    
+\frac{1}{2}\Phi^TV_\mu^T\partial_\mu\Phi+\frac{1}{2}\partial_\mu\Phi^TV_\mu\Phi \nonumber \\
& + & \frac{1}{2}\Phi^TV^T_\mu V^\mu\Phi    
+\frac{1}{4}(W_{\mu\nu}^i)^2+\frac{1}{4}(B_{\mu\nu})^2]   \nonumber
\eaa
 with:
 \ber
  \sqrt{g}d^4x=e^{2ax}dt_Edxdydz
 \eer
    and the integrals are performed on the region $t_E \in [0,2 \pi /a]$  and $x,y,z \in (-\infty,\infty)$.

 As commented above we are interested in making  the computations of the partition function in the large $N$ limit. This limit makes sense if  we take also the limit $\lambda, g^2$ and $g'^2$ going to zero with $N \lambda, N  g^2$ and $N g'^2$ constant. To implement these limits a standard technique consists (see for example \cite{Coleman}) of introducing an auxiliary scalar field $\chi$ so that
\ber
Z_R(a)  = \int[dW][dB][d\chi][d \sigma] [d  \bar \pi] \exp
\left( -\tilde S_{RE}[\bar \pi, \sigma,\chi,W,B] \right)   \nonumber
\eer 
with
\baa
\tilde S_{RE}[\bar \pi,\sigma,\chi,W,B]  & = & \int d^4x \sqrt{g}  [
\frac{1}{2}\pi^a(-\square_E  + \chi)\pi^a    \nonumber   \\
& + & \frac{1}{2}\sigma(-\square_E  )\sigma    
+\frac{1}{2}(\sigma^2 - v^2)\chi  \nonumber  \\
 & -  & \frac{\chi^2}{16 \lambda}-\lambda v^4 + ...]
\eaa
where we have omitted the terms involving gauge fields which are not $\chi$ dependent. By integrating this field,  which is not dynamical, one can immediately recover the previous partition function. In fact the (algebraic) Euler-Lagrange equation for $\chi$ reads:

\ber
\chi =  4 \lambda (\bar \pi^2 + \sigma^2-v^2).
\eer

Now we can perform a standard Gaussian integration of the $\pi^a$ fields and we
get:
\ber
e^{ -   \Delta\Gamma[\chi] }=
\int [d\bar \pi ] e^{- \frac{1}{2}\int d^4x \ \sqrt{g} \pi^a
\left[ - \square_E + \chi \right] \pi^a }
\eer
with:
\ber
\Delta\Gamma[\chi] = \frac{N}{2} Tr
\log\frac{-\square_E+\chi}{-\square_E}.
\eer
Thus we have:
\baa
Z_R(a)=\int [dW][dB]  [d\chi] [d\sigma]e^{-\Gamma_R[\sigma,\chi, W, B]}
\eaa
where the effective action in the exponent is:
\baa
\Gamma_R[\sigma,\chi,W,B] & = & \int d^4 x \sqrt{g}[
\frac{1}{2} \sigma \left( - \square_E \right) \sigma
+ \frac{1}{2 }\left( \sigma^2-v^2 \right) \chi       \nonumber   \\
   & -  & \frac{\chi^2}{16\lambda}-\lambda v^4+\frac{N}{2}\log\frac{-\square_E+\chi}{-\square_E}     \nonumber  \\
   & + &  \sigma^2(\frac{g^2}{8} W^2 +\frac{g'^2}{8} B^2 +\frac{1}{4}g g'  W^3_\mu B^\mu     )      \nonumber   \\
   & + &  \frac{1}{4}(\tilde W_{\mu\nu}^i)^2+\frac{1}{4}(B_{\mu\nu})^2] +O(1/N)  
   \eaa
where 
\ber
\tilde W^i_{\mu\nu}=\partial_\mu W_\nu^i-\partial_\nu W_\mu^i.
\eer
 Notice that the explicit terms in the first two lines are order $N$ and the ones in the third and fourth lines are order one in the large $N$ limit considered here. The quadratic terms in the gauge fields can be diagonalized as usual by introducing the fields
 \ber 
W_\mu^\pm = \frac{1}{\sqrt{2}}(W_\mu^1 \mp i W_\mu^2)
\eer
and
\ber
Z_\mu = \cos \theta _W W_\mu^3 - \sin \theta_W B_\mu
\eer
where $\theta_W$ is the Weinberg angle with $\tan \theta_W = g'/g$.
The orthogonal combination is the photon field:
\ber
A_\mu = \sin \theta _W W_\mu^3 + \cos \theta_W B_\mu.
\eer
but this field does not appear in the quadratic terms.
Obviously these terms will produce masses for the $W$ and $Z$ electroweak bosons whenever the field $\sigma$ (in fact $\sigma^2$) develops a VEV.
 
 The functional integral above can be computed in the large $N$ limit by expanding the fields around some point in the functional space $\overline{\sigma}, \overline{\chi}, \overline{W}$ and $\overline{B}$
 where the first derivative of $\Gamma_R[\sigma,\chi,W,B]$ vanishes. Then, by using the steepest descent method one has
  \ber
Z_R(a)=e^{-\Gamma_R[\overline{\sigma}, \overline{\chi}, \overline{W},\overline{B}]}+O(1/\sqrt{N}), \nonumber
\eer
where we have taken into account that $\Gamma_R[\sigma,\chi]$ is
order $N$. Then, in the large $N$ limit  we have:
\ber
\overline{\sigma}^2(x)    =  ( < \Omega_M \mid\hat \sigma (x) \mid \Omega_M>)^2= < \Omega_M \mid (\hat \sigma (x) )^2 \mid \Omega_M>  \nonumber
\eer
and therefore the $W^{\pm}$ and $Z$ masses will be given by:
\ber
M_W^2=\frac{1}{4}g^2 \overline{\sigma}^2(x) 
\eer
and
\ber
M_Z^2=\frac{1}{4}(g^2+g'^2) \overline{\sigma}^2(x). 
\eer
Notice that in general  the masses are position dependent because  they are produced by the Higgs mechanism  in a Rindler space which is not homogeneous.

Now we can choose     $\overline{W} =\overline{B}=0$   and   $\overline{\sigma}$ and $\overline{\chi}$ as
the solutions of:
 \baa
\frac{\delta \Gamma_R}{\delta \sigma(x)}   &  = & - \square_E\sigma
+ \chi \sigma =0 \label{Equ1} \\
\frac{\delta \Gamma_R}{\delta \chi(x)} &  =  & \frac{1}{2}\left(
\sigma^2 -v^2 \right)      \nonumber   \\   & - & \frac{\chi}{8
\lambda}+
\frac{N}{2}  G(x,x;\chi) \label{Equ2}
\eaa
where
\begin{equation}
\left( - \square_E + \chi \right)_xG(x,x';\chi)=
\frac{1}{\sqrt{g}}\delta^{(4)}(x-x')
\end{equation}
with the boundary conditions $\overline{\sigma}=v$ and $\overline{\chi}=0$ in the limit $x$ going to infinity.

\section{The acceleration driven phase transition}
 
 In principle the above equations are very difficult to solve for $x$ depending fields. However we can proceed in a similar way as in \cite{Dobado} where the above    equations  were considered  in the context of the (not gauged) $SO(N+1)/SO(N)$ LSM. The result goes as follows (see Appendix A). At $x=0$, i.e., the origin of the comoving frame with acceleration $a$, there are two possible solutions depending on the $a$ value. If $a$ is not bigger than the critical acceleration $a_c$ given by 
  \ber
  a_c= 4 \pi v \sqrt{\frac{3}{N}}= 4 \sqrt{3}
\pi F \ee
$(0<a<a_c)$ then:
\ber
\overline {\sigma} (0)=  v  \sqrt{1-\frac{a^2}{a_c^2}}
\eer
and $\overline{\chi}=0$.  However, for $a>a_c$ we have
 \ber
\overline {\sigma} (0)= 0
\eer
 and $\overline{\chi}$ different from zero. Notice that the critical acceleration $a_c$ is $N$ independent. These two cases are associated with two different phases of the system. The first one is the broken phase where we have SSB of the $SU(2)_L \times U(1)_Y$ gauge symmetry  and the electroweak gauge bosons have masses given by: 
 \ber
M_W^2=\frac{1}{4}g^2 v^2(1-\frac{a^2}{a_c^2})
\eer
and
\ber
M_Z^2=\frac{1}{4}(g^2+g'^2) v^2(1-\frac{a^2}{a_c^2}).
\eer
In the second phase ($a> a_c$)  we have a restoration of the $SU(2)_L \times U(1)_Y$ gauge symmetry and consequently:
\ber
M_W^2=M_Z^2=0.
\eer
This is a typical second order Ginzburg-Landau phase transition but with the acceleration playing the role of the temperature. Therefore the accelerating observer experiments a phase transition (restoration of the electroweak gauge symmetry of the SM) at the critical acceleration:
  \ber
  a_c= 4 \pi v \simeq 3 \   TeV
 \eer
 for $N=3$. Notice however that, as commented above, $a_c$ is formally $N$ independent since $v$ is of the order of $\sqrt{N}$.

\begin{figure}[tb]
  \begin{center}
    \includegraphics[width=0.4\textwidth]{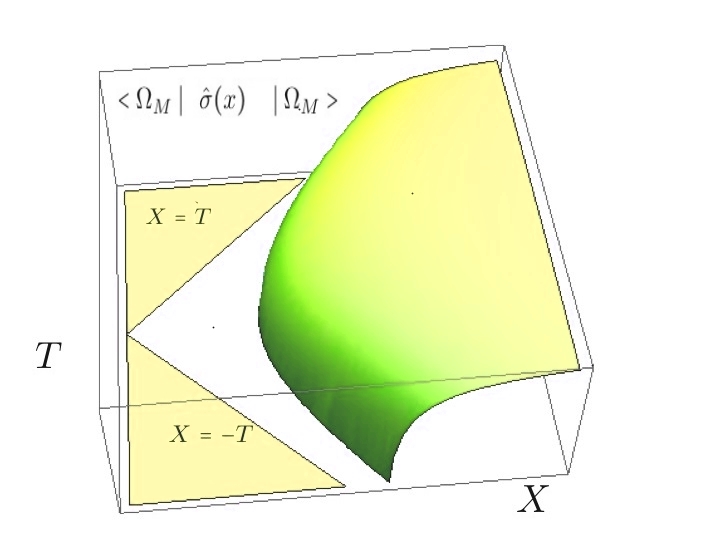}
    \caption{\label{fig:landscape} Profile of the Higgs VEV  for different points of the space-time as seen by the accelerating observer}
  \end{center}
\end{figure}

Next it is interesting to consider what happens at points in Rindler space with $x$  different from cero. Equivalently we can consider  
 a different accelerating observer at Rindler
coordinate $\rho' = 1/a'$. This observer will find a similar
result just exchanging $a$ by $a'$. From the point of view of the
first observer the second observer is located at some point $x$
given by:
\ber
\rho'=\frac{1}{a'}=\frac{1}{a}e^{a x}=\frac{1}{a(x)},
\eer
i.e., the acceleration of the second observer is $a'=a(x)=a e^{-ax}$. Now  it is immediate to find the position dependent
 squared VEV of the $\sigma$ field which, in comoving coordinates, is given by:
\baa
\bar \sigma^2(x)  &  =  & < \Omega_M \mid   (\hat \sigma (x) )^2 \mid \Omega_M> \nonumber \\
&  =  & v^2\left( 1- \frac{a^2}{a_c^2}e^{-2ax} \right)
\label{result}
\eaa
which implies $x$ dependent electroweak bosons masses:
 \ber
M_W^2=\frac{1}{4}g^2 v^2\left( 1- \frac{a^2}{a_c^2}e^{-2ax} \right)
\eer
and
\ber
M_Z^2=\frac{1}{4}(g^2+g'^2) v^2\left( 1- \frac{a^2}{a_c^2}e^{-2ax} \right).
\eer 
For  a comoving frame, with 
acceleration $a$ belonging to the interval $0 < a < a_c$, the electroweak gauge boson masses are a function on the coordinate $x$ ranging
from the standard value in Minkowski space ($a=0$) for $x=\infty$ to zero at the critical value $x_c$ given by:
\ber
x_c=\frac{1}{a}\log \frac{a}{a_c  } < 0
\eer
At this this point (in fact a surface because of the transverse coordinates $x$ and $y$), the phase transition takes place and the $SU(2)_L \times U(1)_Y$ gauge symmetry is restored. Of course the symmetry is also restored on  the region close to the horizon $x< x_c$. In Rindler coordinates one has symmetry restoration in the region defined by $\rho < 1/a_c$. Thus the accelerating trajectory with acceleration $a_c$ defines the boundary between the regions corresponding to the two different phases. The position of this boundary depends only on $v$ but not on the acceleration $a$. Therefore the landscape of the VEV in Rindler space depends only on the parameters of the SM but not on any other acceleration but the critical one. In terms
of the Minkowski coordinates $X$ and $T$ the VEV in the broken phase is given by:
\ber
\bar \sigma^2=v^2\left( 1- \frac{1}{ a_c^2\rho^2}   \label{VEV1}
\right)=v^2\left( 1- \frac{1}{ a_c^2(X^2-T^2)} \right).
\eer
which is plotted  in Fig.~\ref{fig:landscape}. 

\vspace{1cm}

\section{Discussion}    

The possibility of having symmetry restoration by acceleration, as the one considered in this paper, has been sometimes considered at least controversial because of the following argument (see for example  \cite{Hill:1985wi}). Let $\sigma_M(X)$ be the corresponding classical $\sigma$ field in Minkowski (inertial) coordinates collectively denoted by $X$. Then, as $\sigma$ is an scalar, one should have on the right wedge $\cal R$:
\begin{equation}
\sigma(x)= \sigma_M(X)   \label{scalar}
\end{equation}

On the other hand the VEV of the Minkowski  quantum field $\hat \sigma_M(X)$ is given by:
\begin{equation}
< \Omega_M \mid   \hat \sigma_M (X)  \mid \Omega_M > =v. \label{VEVI}
\end{equation}
since the symmetry is spontaneously broken for inertial observers. Is this not in contradiction with the result:
\begin{equation}
< \Omega_M \mid   \hat \sigma (x) \mid \Omega_M> =  v \sqrt{ 1- \frac{a^2}{a_c^2}e^{-2ax} }  \label{VEVII}
\end{equation}
found in this work?
 The answer clearly is not, since Eq.\.(\ref{scalar}) does not imply that 
$< \Omega_M \mid  \hat \sigma_M (X)  \mid \Omega_M>  $  has to equal $< \Omega_M \mid   \hat \sigma (x)  \mid \Omega_M>$.
The reason is the following: the Minkowski Hilbert space can be split as ${\cal H}_M={\cal H}_L  \otimes {\cal H}_R$, where
${\cal H}_L $ and  $ {\cal H}_R $ are the Hilbert spaces corresponding to the regions ${\cal L}$  and ${\cal R}$ respectively.
$\hat \sigma_M (X)$ is an operator defined on the whole Minkowski Hilbert space ${\cal H}_M$. However 
 $\hat \sigma_R (x)=\hat \sigma (x) $ is an operator defined only on  ${\cal  H}_R$, and it must be understood as $1 \otimes \hat \sigma_R (x) $ when acting on $ \mid \Omega_M > \in  {\cal H}_M$. Events belonging to the region ${\cal P}$ can affect events both in ${\cal L}$ and ${\cal R}$.  Therefore the field quantum fluctuations in both wedges are entangled and this means that $\hat \sigma_M$ is not the tensorial product of 
$\hat \sigma_L$ and $\hat \sigma_R$  i.e.:
\begin{equation}
\hat \sigma_M(X) \neq \theta(-X)\hat \sigma_L(x)  \otimes  \theta(X)\hat \sigma_R(x).
\end{equation}
As a consequence  Eq.\.(\ref{VEVI}) and Eq.\.(\ref{VEVII}) are not incompatible at all.

Another important point concerning our results is the following. Introducing the Unruh like critical temperature:
\begin{equation}
T_c= \frac{a_c}{2 \pi}= 2v
\end{equation}
and
\begin{equation}
T(x)= \frac{a}{2 \pi}e^{-ax}= \frac{1}{2\pi \rho}
\end{equation}
the VEV in the comoving frame is given by:
\begin{equation}
< \Omega_M \mid   \hat \sigma (x) \mid \Omega_M> =  v \sqrt{ 1- \frac{T(x)^2}{T_c^2}} .
\end{equation}
In other words it is like if the Higgs field were feeling a thermal bath with a space-dependent temperature $T(x)$    \cite{Candelas:1977zza} which diverges at the horizon and goes to zero at the infinite. Notice that this is compatible with the Tolman and Ehrenfest   \cite{Tolman:1930ona} rule for thermal equilibrium in static space-times since:
\begin{equation} 
T(x)\sqrt{g_{00}}= T(0)e^{-ax}e^{ax}=\frac{a}{2\pi}
\end{equation}
is a $x$ independent constant.

The critical acceleration we have found for the restoration of electroweak symmetry is very large, $a_c \simeq 3$ TeV, which is $1.35 \times 10^{36}$ m/s$^2$, i.e. 35 orders of magnitud larger than the acceleration of gravity on Earth. On the other hand, according to our previous discussion, that means that the phase transition occurs at a distance  $\rho \sim 1/a_c$ of the horizon which is approximately $0.66\times 10 ^{-4}$ fm.  This is indeed a very small distance and it is difficult to figure out a physical scenario where the phenomenon of electroweak symmetry restoration could take place. However, as commented at the introduction, it is interesting to mention that the LHC is currently studying proton-proton collisions at a center of mass energy of $13$ TeV, which corresponds typically to parton-parton collisions at several TeV's. Thus it is not discarded that the electroweak symmetry restoration by acceleration considered here could play a role in this kind 
of processes. Of course a much more detailed analysis is needed but in any case  we understand that this physical effect is still interesting from the fundamental point of view.

\section{Conclusions}

To conclude it is possible to say that the Thermalization Theorem (the Unruh effect) applies to any interacting  (not only free) QFT  with any kind of fields (scalar, fermionic, gauge, etc) \cite{Lee}. The Unruh temperature $T= a / 2 \pi$ is not just a formal artifact but it is a real temperature which  can give rise to collective non-trivial phenomena such as phase transitions and symmetry restorations. In particular in this work we have shown how the Unruh effect can produce a restoration of the electroweak $SU(2)_L \times U(1)_Y$ gauge symmetry of the the SM (inverse Higgs effect). This means that for an accelerating observer the symmetry is restored for accelerations bigger than a critical value $a_c = 4 \pi v\simeq 3 $ TeV. For $a>a_c$ the electroweak gauge bosons become massless as the photon. Also we have seen that for such an accelerated observer with $a<a_c$, the symmetry is also restored beyond a surface defined by $ x<x_c= \log (a/a_c)/a$ (in the horizon direction),  where the electroweak gauge bosons are massless. In fact this happens also to any other elementary particle (quarks, leptons and the Higgs boson itself) since in the SM all of them have masses controlled by the Higgs VEV. As a consequence all (elementary) particles become massless for enough accelerated observers. We think this is a very interesting result at the fundamental level, coming from  the formulation of the SM as a QFT on Rindler space-time. In addition there are some possibilities that it could play a role at the LHC or other higher energy colliders in the future.

\subsection{ACKNOWLEDGMENTS}

The author thanks E. \'Alvarez and  R. Tarrach for triggering our interest in the problem considered in this work,  L. \'Alvarez-Gaum\'e for comments concerning \cite{Bell}, C. Pajares for bringing into our attention references \cite{Kharzeev:2005iz} and \cite{DiasdeDeus:2006xk} and for  useful discussions and J. A. Ruiz-Cembranos for reading the manuscript. Work supported by Spanish grants MINECO:FPA2014-53375-C2-1-P  and FPA2016-75654-C2-1-P.

\subsection{APPENDIX A}

Here we will find approximate solutions for  the equations  Eq.~(\ref{Equ1}) and  Eq.~(\ref{Equ2}) to obtain $\bar \sigma$ and $\bar \chi$.
In particular we  consider the region $ax<<1$. In this regime the
accelerating observer goes into the Minkowski inertial frame for fixed $x$ ($a$ goes to zero) or $x$ goes to zero for fixed $a$. Thus we look for solutions with vanishing $\overline{\chi}$. In Appendix B it is shown how in this case our equations become:
  \begin{eqnarray}
0  & = &    \square_E\sigma     \\
0 & = & \sigma^2 -v^2 +\frac{N}{2\pi^3}\int_0^\infty d \Omega 
\frac{\Omega \pi}{2 \rho^2 \tanh(\Omega \pi)}.
   \label{Equ2p2}
\end{eqnarray}

Introducing $\omega$ as $\omega=a \Omega$ and using $\rho a = 1+ a
x+...$ we find, up to order $a x$:
\begin{equation}
\sigma^2=v^2 -\frac{N}{4\pi^2}(1-2ax)\int _0^\infty d \omega
\omega\left( 1 + \frac{2}{e^{\frac{2 \pi}{a}\omega}-1} \right)+... \nonumber
\end{equation}
Obviously the first integral requires some regularization and renormalization.  This
can be done by using a $x$ dependent ultraviolet
cutoff $\Lambda e^{-ax}$  and performing the renormalization
of the $v$  parameter:
\begin{equation}
v^2 \rightarrow   v^2-N\frac{\Lambda^2}{2 (2 \pi)^2}(1-2ax+...).
\end{equation}
This renormalization naturally matches the  $a=0$ limit and is consistent with  the red/blue shift detected by the accelerating observer
when receiving a signal emitted at the point $x$. Then we have:
\begin{equation}
\sigma^2=v^2 -\frac{N}{2\pi^2}(1-2ax)\int _0^\infty d
\omega\omega\frac{ 1}{e^{\frac{2 \pi}{a}\omega}-1} +O((ax)^2) .   \nonumber
\end{equation}
By performing the  $\omega $ integration, the Minkowski VEV of the $\hat \sigma^2(x) $ comoving
operator is given in the $ax<<1$ regime by:
\begin{eqnarray}
\bar
\sigma^2(x)  & = & < \Omega_M \mid (\hat \sigma (x) )^2 \mid \Omega_M>   \\
   & =  & v^2\left( 1- \frac{a^2N}{12 (2\pi)^2 v^2}(1-2ax)   \right).     \nonumber
 \end{eqnarray}
By introducing the critical acceleration:
\begin{equation}
a_c^2= 3(4 \pi)^2 \frac{v^2}{N}
\end{equation}
we have:
 \begin{equation}
\bar \sigma^2(x)=v^2\left( 1- \frac{a^2}{a_c^2}+
2\frac{a^3}{a_c^2}x+... \right) \label{linear}.
 \end{equation}
Notice that at this order this is also a solution of
Eq.~(\ref{Equ1}). Therefore, at the origin of the accelerating    
frame ($x=0$ or $\rho=1/a$), the squared VEV of the $\hat \sigma$
field is given by:
\begin{equation}
\bar \sigma^2(0)=< \Omega_M \mid (\hat \sigma (0) )^2 \mid \Omega_M>=v^2\left(
1- \frac{a^2}{a_c^2} \right)
 \end{equation}
 for $0 \le a \le a_c$ and clearly:  
   \begin{equation}
< \Omega_M \mid   (\hat \sigma (0) )^2 \mid \Omega_M>=0
 \end{equation}
for $a > a_c$. This is exactly the thermal behavior of the LSM
in the large $N$ limit with $a/a_c$ playing the role of $T/T_c$
(as seen by a  inertial observer).
It corresponds to a second order phase transition at the critical
acceleration $a_c$ where the original spontaneously broken
symmetry is restored for the accelerating observer.

\subsection{APPENDIX  B}
Here we will give some of the details on the computation of the
Euclidean Green function $G(x,x';s)$ defined by:
\begin{equation}
\left( - \square_ E \right)_xG(x,x';s)=
\frac{1}{\sqrt{g}}\delta^{(4)}(x-x')
\end{equation}
for constant $s$ and the appropriate boundary conditions which are periodic in
the time coordinate with periodicity $\beta=1/T=
2\pi/a$ . We can use Rindler
coordinates where:
\begin{equation}
\square_E=\partial_\rho^2+\frac{1}{\rho}\partial_\rho+\frac{1}{\rho^2}\partial_\eta^2+\partial_\bot^2.
\end{equation}
Now we introduce the partial Fourier transform:
\begin{eqnarray}
& (2\pi)^2 &G(\rho,\rho',x_\bot-x'_\bot,\eta-\eta';s)  \nonumber   \\
 & = & \sum_{n=-\infty}^\infty\int
dk_\bot^2 e^{i[n(\eta-\eta')+k_\bot(x_\bot-x'_\bot)]}  \tilde
   G(\rho,\rho',k_\bot,n;s)     \nonumber
\end{eqnarray}
which satisfies:
\begin{equation}
[\rho^2\partial_\rho^2+\rho\partial_\rho-(\alpha^2\rho^2+n^2)]\tilde
G= -\rho\delta(\rho-\rho')
\end{equation}
where $\alpha^2=k^2_\bot+s$. The solution can be written as:
\begin{equation}
\tilde G(\rho,\rho',k_\bot,n;s)=\int_0^{\infty} d\Omega
\frac{\Psi_\Omega(\rho)\Psi_\Omega(\rho')}{\Omega^2+n^2}
\end{equation}
where $\Psi_\Omega(\rho)$ can be obtained from the solution
of the modified Bessel functions with imaginary parameter:
\begin{equation}
\Psi_\Omega(\rho)=\frac{1}{\pi}\sqrt{2 \Omega \sinh (\Omega
\pi)}K_{i\Omega}(\alpha\rho).
\end{equation}
By using well known properties of these functions and:
\begin{equation}
\sum_{n=-\infty}^{\infty}\frac{1}{\Omega^2+n^2}=\frac{\pi}{\Omega}\frac{1}{\tanh(\Omega
\pi)}
\end{equation}
it is possible to find:
\begin{equation}
G(x,x;s)=\frac{1}{2 \pi^3} \int_0^\infty d \Omega\cosh(\Omega
\pi)\int_0^{\infty}dk k K_{i \Omega}^2(\alpha\rho).  \nonumber
\end{equation}
Now taking $s=0$   it is straightforward to get  Eq.~(\ref{Equ2p2}).\\

\end{document}